\documentclass[aps,prl,twocolumn,superscriptaddress,longbibliography]{revtex4-1}
\usepackage{graphicx}
\usepackage{amsmath}
\usepackage{url}
\usepackage{xcolor}

\begin{document}
	
	%\preprint{}
	
	\title{Controlled and robust two-mode emission from the interplay of driving and thermalization in a dye-filled photonic cavity}
	
	
	\author{M. Vlaho}
	\email[]{vlaho@pks.mpg.de}
	%\thanks{}
	\affiliation{Max-Planck-Institut f{\"u}r Physik komplexer Systeme, N{\"o}thnitzer Straße 38, 01187 Dresden, Germany}
	
	\author{H.A.M. Leymann}
	\email[]{ham.leymann@gmail.com}
	\affiliation{Max-Planck-Institut f{\"u}r Physik komplexer Systeme, N{\"o}thnitzer Straße 38, 01187 Dresden, Germany}
	\affiliation{INO-CNR BEC Center and Dipartimento di Fisica, Universita di Trento, I-38123 Povo, Italy}
	
	\author{D. Vorberg}
	%\email[]{}
	\affiliation{Max-Planck-Institut f{\"u}r Physik komplexer Systeme, N{\"o}thnitzer Straße 38, 01187 Dresden, Germany}
	
	\author{A. Eckardt}
	\email[]{eckardt@pks.mpg.de}
	\affiliation{Max-Planck-Institut f{\"u}r Physik komplexer Systeme, N{\"o}thnitzer Straße 38, 01187 Dresden, Germany}
	
	
	\date{\today}
	
\begin{abstract}
Two dimensional photon gases trapped in dye-filled microcavities can undergo thermalization and nearly ideal equilibrium Bose-Einstein condensation.  
However, they are inherently driven-dissipative systems that can exhibit an intricate interplay between the thermalizing influence of the environment given by the dye solution and the pump and loss processes driving the system out of equilibrium.
We show that this interplay gives rise to a robust mechanism for two-mode emission, when the system is driven by an off-centered pump beam.
Namely, after the system starts lasing in the dominantly pumped excited mode, in a second transition a photon condensate is formed in the ground mode, when the pump power is increased further. This effect is a consequence of the redistribution of excited dye molecules via the lasing mode in combination with thermalization. We propose to exploit this effect for engineering controlled two-mode emission and demonstrate that by tailoring the transverse potential landscape for the photons, the threshold pump power can be tuned by orders of magnitude.
% We consider a situation, where this competition is enhanced by an off-centered pump spot not overlapping with the ground mode. We find that under such conditions the system undergoes two transitions when the pump-power is ramped up: At a first threshold, the system starts to lase in the dominantly pumped excited mode. The overlap of the largely occupied lasing mode with the ground mode, then triggers a second transition, where thermalization leads to the formation of an additional photon condensate in the ground mode. This mechanism can provide a robust tool for engineering devices with controlled multi-mode emission. We, moreover, show that by employing a recently developed experimental tool for tailoring the transverse potential landscape for the photons, the second threshold can be tuned by orders of magnitude by tuning resonances between excited cavity modes.
\end{abstract}

%\keywords{}
	
\maketitle

%\paragraph{Introduction}
%After the realization of Bose-Einstein condensation (BEC) of polaritons, coherent superpositions of photons and excitons, \cite{deng_condensation_2002,kasprzak_bose-einstein_2006,byrnes_excitonpolariton_2014,balili_bose-einstein_2007,plumhof_room-temperature_2014,sun_bose-einstein_2017,sun_observation_2012}, thermalization \cite{klaers_thermalization_2010} and ultimately the formation of a BEC of photons \cite{klaers_boseeinstein_2010,marelic_experimental_2015} has been experimentally realized.
A system of photons in a dye-filled cavity can be used as a platform for studying the interplay between gain and loss on the one hand and thermalization (via the rovibrational relaxation of the dye molecules interacting with the environment given by the solvent) on the other.
While the former process describes simple lasing \cite{siegman_lasers_1986}, in the last decade the regime where thermalization is the dominant process, has been realized and equilibrium-like Bose condensation of photons was observed in various systems \cite{klaers_boseeinstein_2010, marelic_experimental_2015,weill_boseeinstein_2019,rajan_photon_2016,nyman_bose-einstein_2018,walker_driven-dissipative_2018}. Also the temporal \cite{hesten_collective_2018,hesten_non-critical_2018,schmitt_dynamics_2018} and spatial \cite{keeling_spatial_2016} features of photon BECs driven out of equilibrium have been studied, as well as the demarcation of photon BECs from lasers \cite{schmitt_thermalization_2015,schmitt_bose-einstein_2016,leymann_pump-power-driven_2017,vorberg2018unified,radonjic_interplay_2018}, the grand-canonical statistics in photon BECs \cite{schmitt_observation_2014}, the breakdown of equilibrium-like behavior \cite{kirton_thermalization_2015} and the thermo-optic interaction effects \cite{stein2019collective}. Very recently, it was found both theoretically 
\cite{hesten_decondensation_2018} and experimentally \cite{walker_driven-dissipative_2018} that excited cavity modes start to emit coherently together with the ground mode, when the pump power and the photon loss are increased relative to the thermalizing coupling to the dye. A complex network of small patches in parameter space corresponding to various phases characterized by different combinations of modes with macroscopic occupation is predicted \cite{hesten_decondensation_2018}. 

While the last described situation gives rise to rich physics, it is not an ideal starting point for the robust engineering and the control of multi-mode emission. 
In this paper we, therefore, explore an alternative non-equilibrium scenario, where, in contrast to previous studies, the interplay between driving and thermalization is controlled by an \emph{off-centered} pump beam.  
We find that the system undergoes two pump-power driven non-equilibrium phase transitions. First, the system starts to lase in an excited mode, which is directly determined by the position of the pump spot. 
When the pump power is increased further, the spatial redistribution of pump power mediated by this lasing mode then triggers a second transition, where thermalization leads to the additional formation of an equilibrium-like Bose condensate in the ground mode. 
In a system where both drive and thermalization are present, a sharp distinction between lasing and Bose condensation is, strictly speaking, no longer possible. 
Nevertheless, the characterizations of the first transition as lasing and the second as condensation provides a useful way to mark the mechanisms (selective pumping vs.~thermalization) that are mainly responsible for the mode selection. 
The fact that the lasing mode can be selected by adjusting the pump spot, while the second transition always corresponds to the on-set of ground-state condensation, makes this mechanism of lasing-assisted Bose condensation a promising tool for engineering systems with robust and tunable two-mode emission. 
In order to explore this prospect further, we investigate how far the effect can be controlled by shaping the transverse potential landscape in the cavity, as it can be done by using recently developed experimental tools based on thermo-optic imprinting \cite{dung_variable_2017}.
When pumping the upper minimum of an asymmetric double well, the second transition threshold can be shifted by orders of magnitude by tuning the system close to or further away from interwell resonances. 
%Thus, compared to the very complex phase diagram for multi-mode condensation found in a symmetrically pumped harmonic potential \cite{hesten_decondensation_2018}, the asymmetrically structured and pumped system provides a highly controllable and robust scenario for engineering and switching two-mode emission.
	
%The finite photon-cavity lifetime is the result of loss mechanisms (like the mirror losses, or spontaneous emission into continuum modes) where photons leave the system without going back into the particle reservoir (in our case electronic excitations of the dye). Therefore, the system needs to be pumped with an external laser in order to stabilize the average photon number.
%Strictly speaking one has to consider these photon gases always as driven dissipative systems \cite{radonjic_interplay_2018}, however, careful experimental design allows to recover many of the typical features of equilibrium systems \cite{klaers_statistical_2012,schmitt_observation_2014,schmitt_thermalization_2015} for the photon BECs. 
%Non-equllibrium effects in quantum systems are of high conceptual interest \cite{vorberg_generalized_2013} and have  been studied in various systems e.g.~in cold atoms \cite{bernien_probing_2017}, ions \cite{zhang_defect_2017}, and circuit QED \cite{fink_observation_2017,fitzpatrick_observation_2017}.

%\paragraph{Non-equilibrium steady state}
We describe the system in terms of semiclassical rate equations \cite{kirton_nonequilibrium_2013, keeling_spatial_2016, hesten_decondensation_2018} for the photon mode populations $n_{i}$, and the fraction of excited dye molecules at position $ \vec{r} $, $f(\vec{r})$,
%\begin{widetext}
\begin{eqnarray}
\dot{n}_{i} &=&	-\kappa n_{i} + (n_{i}+1) R_{\downarrow}^{i} \rho \, O_i 
											-n_{i}R_{\uparrow}^{i} \rho \, (1-O_i), \\
%\end{equation}
%\begin{equation}\label{}
%\begin{aligned}
\dot{f}(\vec{r})&=& [1-f(\vec{r})](p(\vec{r})+ \sum_{i}R_{\uparrow}^{i}|\psi_{i}(\vec{r})|^{2}n_{i}) 
\nonumber\\\label{dyn}
&& -\,f(\vec{r})[\Gamma + \sum_{i}R_{\downarrow}^{i}|\psi_{i}(\vec{r})|^{2}(n_{i}+1)].
\end{eqnarray}
%\end{widetext}
Here $\Gamma$ is the rate of spontaneous losses into non-cavity modes and $\kappa$ the photon loss rate, which is assumed to be mode independent. The density of the dye molecules is denoted by $\rho$. The transverse photonic modes $\psi_{i}(\vec{r})$ resulting from the two-dimensional (2D) trap imposed by the mirrors correspond to frequencies $\omega_i$. %$p(\vec{r}) =  P \,e^{-(\vec{r} -\mu \hat{x})^2/2\sigma^2}/\mathcal{N} $
The spatially varying pump rate has the form $p(\vec{r}) =  P \,g_{\mu, \sigma}(\vec{r})$, where $ g_{\mu, \sigma}(\vec{r}) $ is a normalized 2D off-centered Gaussian with standard deviation $\sigma$ and mean $\mu \vec{e}_x$. The gain of mode $i$ is quantified by its overlap
$O_i[f(\vec{r})] = \int |\psi_{i}(\vec{r})|^{2} f(\vec{r})\, d\vec{r}$ with the distribution of excited dye molecules $ f(\vec{r}) $.
In a solution, the rovibrational states of the dye molecules relax rapidly to equilibrium. As a result, their occupation numbers need not to be taken into account explicitly and the absorption and emission rates,
$R_{\uparrow}^{i}$ and $R_{\downarrow}^{i}$ satisfy the Kennard-Stepanov law \cite{mccumber1964einstein,klaers_thermalization_2010,moroshkin2014kennard}
$R_{\downarrow}^{i}\propto R_{\uparrow}^{i} \exp{[-\beta \hbar(\omega_{i}-\omega_{z})]}$, where $ \omega_{z} $ denotes the zero-phonon frequency of the dye.

In the following discussion we will use the general term ``Bose selected'' for modes acquiring macroscopic occupation, which subsumes both equilibrium Bose condensation as well as non-equilibrium processes leading to a macroscopic occupation of bosonic modes \cite{vorberg_generalized_2013, leymann_pump-power-driven_2017, vorberg2015nonequilibrium, vorberg2018unified}. The selection of mode $i$ is associated with (approaching) the divergence of the steady-state occupation in that mode, which happens when $O_i$ reaches the threshold value \cite{hesten_decondensation_2018}
\begin{equation}\label{thres}
 O_i^{th} = \frac{R_{\uparrow}^{i}+\kappa / \rho}{R_{\uparrow}^{i}+ R_{\downarrow}^{i}} = \frac{1+ R_{\uparrow}^0 /(R_{\uparrow}^i \xi)}{1+e^{-\beta \hbar(\omega_{i}-\omega_{z})}}.
\end{equation}
Here we have isolated the dimensionless thermalization parameter $\xi=R_{\uparrow}^{0} \rho/\kappa$ \cite{keeling_spatial_2016, hesten_decondensation_2018}, which quantifies the coupling between the photons and the dye relative to the loss. Once a mode is selected, the gain $O_i$ is clamped \cite{siegman_lasers_1986} close to the threshold $O_i^{th}$.

\begin{figure}[t]
	\includegraphics{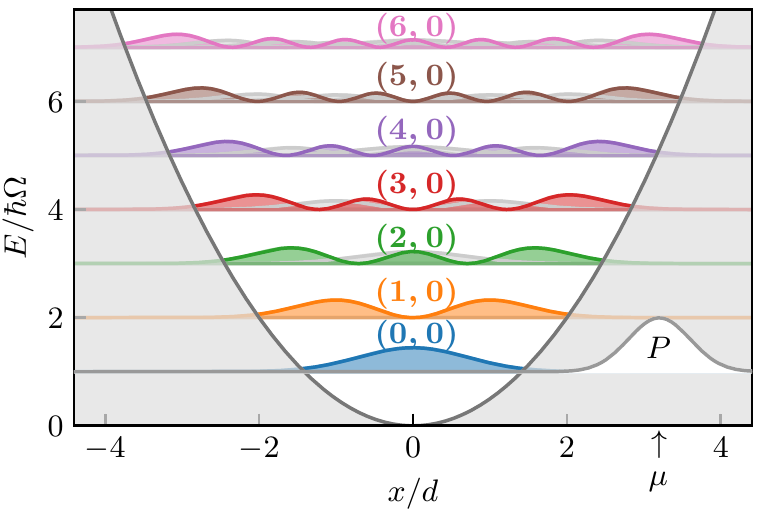}
	\caption{\label{fig:modes} Off-centered Gaussian pump spot ($P$) and photon modes $|\psi_{n_x, n_y}(x)|^{2}$ projected onto the $x$ axis. Modes $(n_x,0)$, with nodes only along $x$ direction, are shown in color.}
\end{figure}

%\paragraph{Harmonic Potential}
We compute the steady state of the system using parameter values corresponding to the experiments of Refs.~\cite{schmitt_thermalization_2015, dung_variable_2017}. We choose room temperature, $T = 300 ~\mathrm{K}$, and a slightly anisotropic harmonic trap, $ V_{HO}(x,y) = (1/2)\,\hbar \, \Omega \, (x^2+Ay^2)/d^2 + V_0$ with $A =1.001$. The frequency spacing is $\Omega/2\pi=2~\mathrm{THz}$, while
$d$ is the harmonic oscillator length. The transverse photonic modes [Fig.~\ref{fig:modes}] are labeled by non-negative integer harmonic oscillator quantum numbers in $x$ and $y$ direction, $i=(n_x,n_y)$. The zero-phonon line is set to 
$\omega_{z}/2\pi = 555\mathrm{THz}$. The frequency of the ground mode, including the longitudinal contribution $\omega_L$, reads $\omega_0 =\omega_L+\Omega= 2 \pi \cdot 510\mathrm{THz}$ and the corresponding absorption rate is %$R_\uparrow(\omega_C) \equiv% 
$R_{\uparrow}^{0} = 1\mathrm{kHz}$. From the measured absorption and fluorescence spectra of the Rhodamine 6G dye \cite{keeling_spatial_2016}, we obtain the corresponding rates $R_{\uparrow, \downarrow}^{i}$ as fitted functions of the frequency $\omega_i$, which satisfy the Kennard-Stepanov law. % Eq.~\eqref{ks}.
The thermalization parameter $\xi$ lies between $0.3$ and $3$, while the 
rate of spontaneous losses into non-cavity modes is set to $\Gamma = 0.2\mathrm{GHz}$. The Gaussian pump spot of width $\sigma=0.4d$ is shifted away from the trap center by $\mu = 3.2d$, so that it has essentially no overlap with the ground mode [see Fig.~\ref{fig:modes}]. In this way, the competition between driving and thermalization is enhanced.  
%Below, we also consider double well traps as they have recently been realized experimentally \cite{dung_variable_2017}.
%We are interested in the interplay between driving via pumping and loss on the one hand and equilibration with the dye solution on the other. Therefore, we consider a situation where the competition between the two processes is enhanced by an off-centered pump spot having essentially no overlap with the ground mode. 
%The modes, which we label by their quantum numbers in $x$ and $y$ direction, $(n_x,\,n_y)$, and the pump profile are depicted in Fig.~\ref{fig:modes}. 
A similar off-centered pump has already been realized experimentally to study the transient relaxation dynamics following a short pump pulse \cite{schmitt_thermalization_2015}. In contrast, we are interested in the steady state of the continuously pumped system.

\begin{figure*}[t]
	\includegraphics[scale=1]{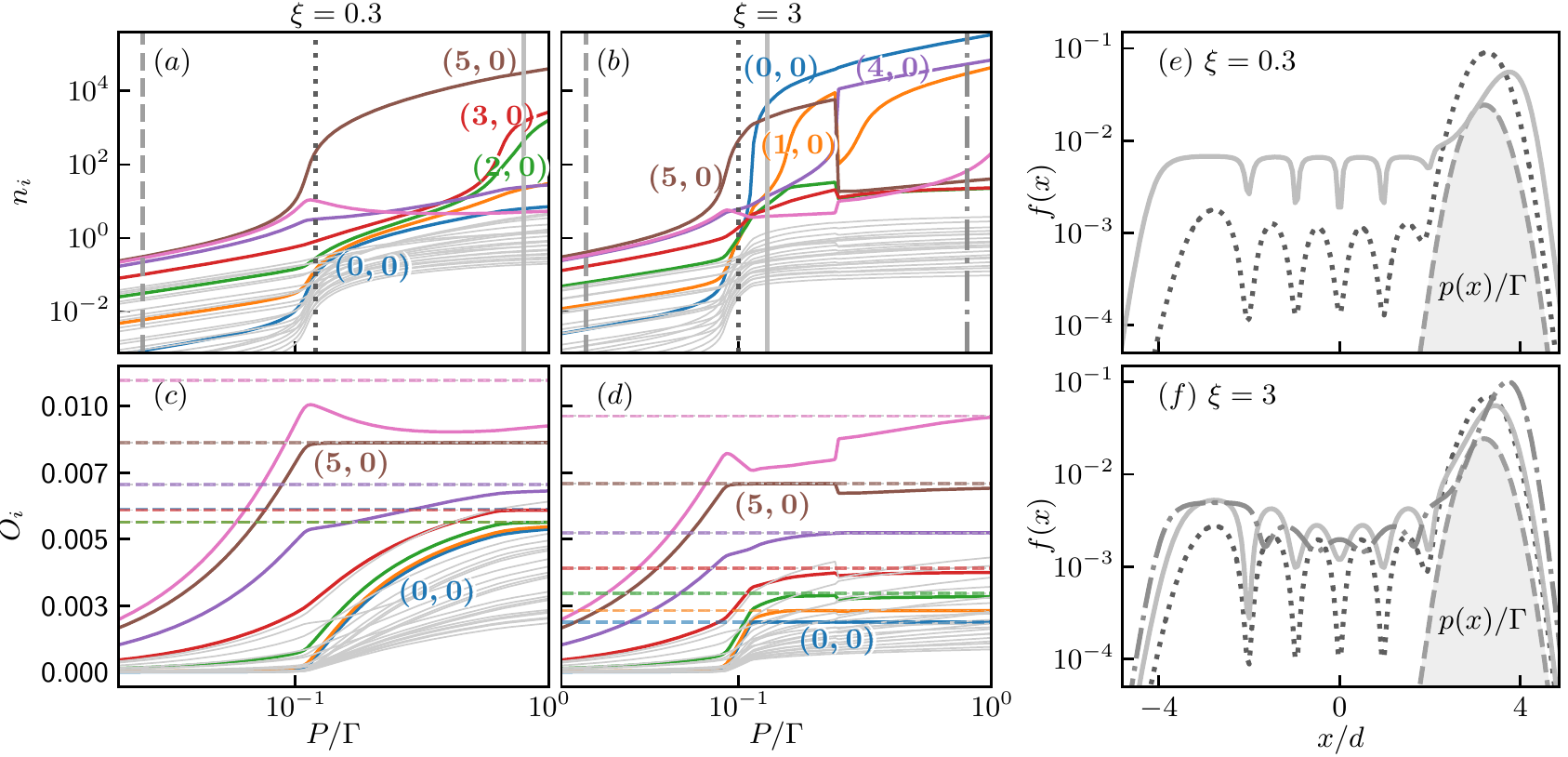}
	\caption{\label{fig:inhomog} Population $n_i$ (a,\,b) and gain $O_i$ (c,\,d) of modes $i$ vs.\ pump rate $P$ for $\xi = 0.3$ (a,\,c) and $\xi=3$ (b,\,d). The dashed line in (c,\,d) indicate $O_i^{th}$. (e,\,f) show the spatial distributions $f(x)$ of excited dye molecules along the $x$-axis for 
	$\xi=0.3,3$ at those $P$ marked by the corresponding vertical lines in (a,\,b),
	respectively. The shaded area represents $p(x)/\Gamma$.}
\end{figure*}

Numerically obtained mode populations for $\xi = 0.3$ and $\xi = 3$ are shown in Figs.~\ref{fig:inhomog}(a) and (b), respectively. The colors correspond to the modes as shown in Fig.~\ref{fig:modes}. In both cases mode (5,0) (brown) is selected first. %When $ \xi $ is sufficiently high (b), the ground mode (blue) also gets selected with further increase of $ P $, followed by the selections and deselections of additional modes \cite{hesten_decondensation_2018}.
Figs.~\ref{fig:inhomog}(c) and (d) depict the corresponding gain $O_i$ of each mode (solid curves) as a function of the pump rate. The threshold values of the gain 
$O_i^{th} $ are shown as the dashed horizontal lines. One can see that each mode selection [Fig.~\ref{fig:inhomog}(a, b)] is accompanied by gain clamping [Fig.~\ref{fig:inhomog}(c,d)].

\begin{figure}[!htbp]
	\includegraphics{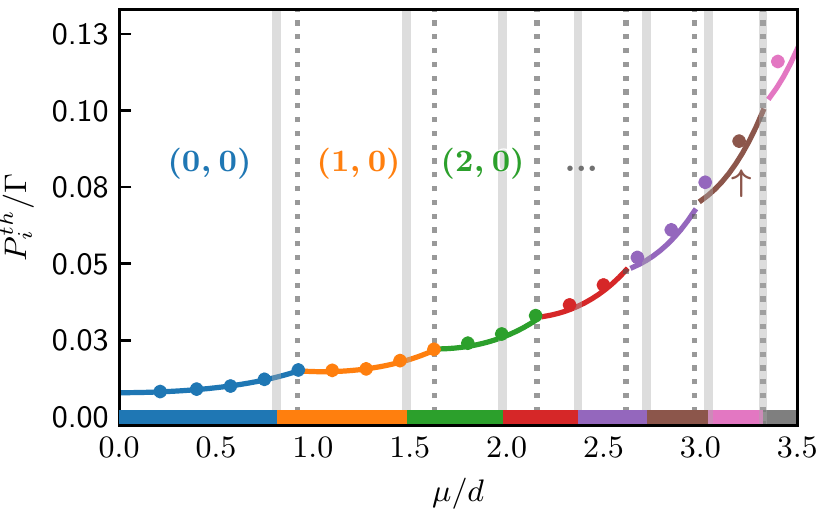}
	\caption{\label{fig:mode_vs_mu} Threshold pump rate $ P_i^{th} $ for the first selection [Eq.\,\eqref{lasing_mode}] vs.\ pump-spot position $\mu$ for
	$\xi=3$. The different colors and labels $(n_x,n_y)$, separated by dotted vertical lines, indicate which mode is selected. The dots are obtained numerically form the full rate equations. The colored bars at the bottom, separated by solid lines, indicate the mode with the largest gain 
	$O_i[g_{\mu, \sigma}(\vec{r})]$.}
\end{figure}

In order to understand, which mode becomes selected first, let us approximate the distribution of excited dye molecules in the steady state below the first threshold by $f(\vec{r}) \approx p(\vec{r})/(p(\vec{r})+\Gamma) \approx p(\vec{r})/\Gamma$. Here the first expression is obtained from Eq.~\eqref{dyn} by neglecting the coupling to the still weakly occupied photonic modes. Inserting this expression into the threshold gain given by Eq.~\eqref{thres}, we get the following condition for the threshold pump rate of mode $ i $
%%%% VERSION 1 %%%%
%\begin{equation}\label{lasing_mode}
%\int \frac{P_i^{th} g(\vec{r}; \mu, \sigma) |\psi_{i}(\vec{r})|^{2}}{P_i^{th} g(\vec{r}; \mu, \sigma) + \Gamma}\, d\vec{r} = \frac{R_{\uparrow}^{i}+\kappa / \rho}{R_{\uparrow}^{i}+ R_{\downarrow}^{i}},
%\end{equation}
%%%% VERSION 2 %%%%
%\begin{equation}\label{lasing_mode}
%\int P_i^{th} g(\vec{r}; \mu, \sigma) |\psi_{i}(\vec{r})|^{2}\, d\vec{r} = \frac{R_{\uparrow}^{i}+\kappa / \rho}{R_{\uparrow}^{i}+ R_{\downarrow}^{i}},
%\end{equation}
%%%% VERSION 3 %%%%
\begin{equation}\label{lasing_mode}
P_i^{th} = \frac{O_i^{th}}{O_i[g_{\mu, \sigma}(\vec{r})]}.
\end{equation}
The selected mode $ i $ is the one with the lowest value of $ P_i^{th} $. We see that there are two competing effects here. While the denominator favors modes having a large overlap with the pump spot (i.e.\ excited modes), the numerator favors modes with low energy. For a narrow pump spot with $\sigma/d\lesssim 1$, as considered here, we expect the former effect to be the dominant one.
Figure~\ref{fig:mode_vs_mu} shows the threshold pump rate $ P_i^{th} $ of the first selection as a function of the pump spot position $\mu$. Results from Eq.~\eqref{lasing_mode} (solid curve) match the exact values obtained numerically (dots) very well. The colors and the labels $(n_x,n_y)$ indicate which mode is selected first; 
it changes at the vertical dotted lines. The colored bars at the bottom, separated by solid vertical lines, in turn indicate the mode with the largest overlap with the pump spot. We can see that the impact of the nominator in Eq.~\eqref{lasing_mode} is to slightly shift the solid lines with respect to the dotted lines. However, as expected, energetics plays a minor role in the selection of the first mode compared to its overlap with the pump spot. For the value $\mu=3.2d$, which was used for the simulations shown in Figs.~\ref{fig:modes} and \ref{fig:inhomog} (arrow in Fig.~\ref{fig:mode_vs_mu}), the selected mode is $(5,0)$, which has only slightly lower gain then the highest gain mode $(6,0)$. Therefore, we refer to the first selection as lasing, since (for $\mu\gtrsim d$) an excited mode is selected predominantly as a consequence of its large gain.

Once the system starts to lase in an excited mode, this mode will create excited dye molecules in an extended region in space, much larger than the narrow off-centered pump spot. This can be seen in Fig.~\ref{fig:inhomog}(e,\,f) showing the spatial distribution $f(x)$ of excited dye molecules along the axis of the pump spot displacement ($x$-axis). The different linestyles correspond to the pump rates indicated by the vertical lines of the same style in Fig.~\ref{fig:inhomog}(a,\,b). As mentioned previously, below the first threshold there are no selected modes and $f(x)\approx p(x)/\Gamma$, as can be seen from the perfect match between the dashed line and the shaded area denoting $ p(x)/\Gamma $. The dotted curve, which shows $ f(x) $ just above the lasing transition, has the additional structure corresponding to the first selected mode. This lasing assisted redistribution of pump-power can then trigger the selection of a second mode. For a sufficiently large thermalization parameter (which lowers the threshold gain $ O_0^{th} $ [Eq.~\eqref{thres}]), this mode is \emph{always} found to be the ground state, which is favored via thermalization with the dye due to its lowest energy. Thus, in this respect, the second transition is akin to equilibrium Bose condensation and we call this effect lasing assisted ground-state condensation. The two facts that (i) the mode which is selected first can be accurately controlled via the position of the pump spot [Fig.~3] and (ii) the second transition always corresponds to the selection of the ground mode, suggest to exploit this effect for controlling two-mode emission in a very robust way.
%At a higher pump rate $P_3$ (dashed curve), the corresponding $f(x)$ has the same structure in Fig.~\ref{fig:inhomog}(e), whereas in Fig.~\ref{fig:inhomog}(f), its shape changes because of the selection of the ground mode.
%The ground mode has a vanishing overlap with the pump spot, and thus no direct access to the pumped gain medium. 
%A higher value of the thermalization parameter ($\xi = 3$) lowers the threshold gain $ O_0^{th} $ [Eq.~\eqref{thres}] resulting in the selection of the ground mode in this case [Fig.~\ref{fig:inhomog}(b,\,d)].
\begin{figure}[!htbp]
	\includegraphics{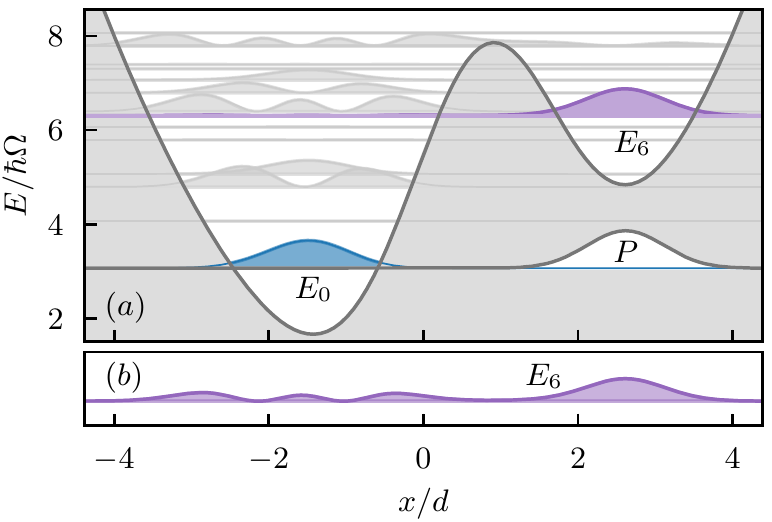}
	\caption{\label{fig:modesDW} (a) Double well potential and modes
	$|\psi_i(x)|^{2}$ for $\delta =0.79d$ together with the pump spot ($P$, white), projected onto the $x$-axis. The modes shown in color (blue and purple) and labeled by their energy are those that get selected. %The dashed vertical line denotes the barrier position $x_B$.
	(b) For a slightly different value $\delta =0.81$, the lasing mode $6$ becomes delocalized between both wells, as a consequence of an interwell resonance. }
\end{figure}
%\paragraph{Double-well potential}

However, in Fig.~\ref{fig:inhomog} (b), we can observe that after the selection of the ground mode in a second transition, further transitions occur. In order to avoid that, and also to have a better control over both the selected modes and their threshold pump rates, let us now consider a structured cavity \cite{dung_variable_2017} imposing a tilted double-well potential for the photons, $V_\text{DW}(x,y)=V_{HO}(x,y) + l \exp{[-(x-\delta)^2/(2\varepsilon^2)]}$. In the following we choose $l=7.5~\hbar\Omega$ and $\varepsilon=1.0~d$, while $\delta$ is used as a tuning parameter. In Fig.~\ref{fig:modesDW}(a) we depict the potential and the corresponding photon modes for $\delta =0.79d$ together with the pump profile $p(\vec{r})$ (white), projected onto the $x$-axis. As in Fig.~\ref{fig:modes}, the modes shown in color are those that get selected.
\begin{figure}[!htbp]
	\includegraphics{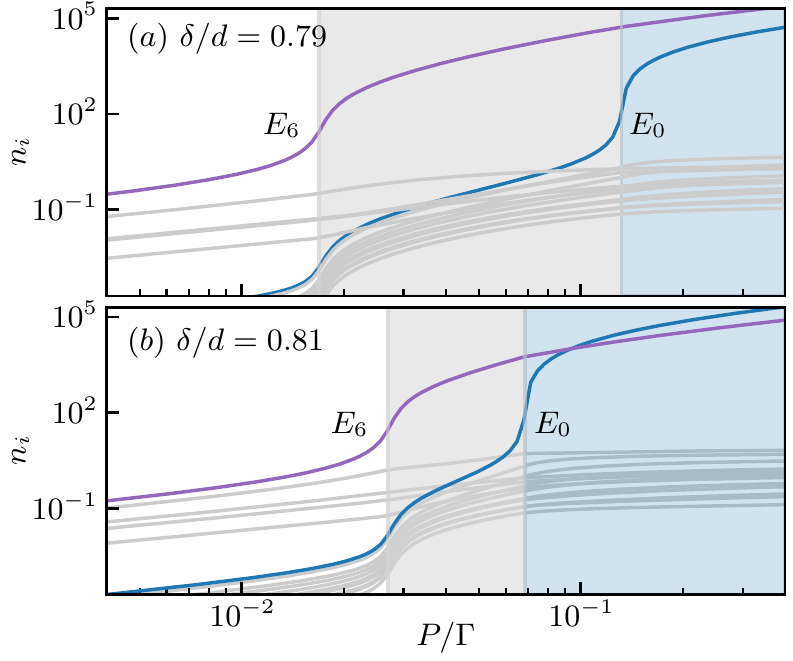}
	\caption{\label{fig:res} Mode populations $n_i$ vs pump rate $P$ for two slightly different values of $\delta$. The grey area indicates a $ P $ range in which there is only lasing, while in the blue region the ground-state condensate is also present.}
\end{figure}

In Fig.~\ref{fig:res}(a) we present the mode populations versus pump power for the parameters of Fig.~\ref{fig:modesDW}(a). The thermalization parameter is 
$\xi=5$, while all the remaining parameters are the same as in the harmonic potential case. Since, essentially, we are only pumping the upper well, mode 6 (purple), having the lowest energy among those modes significantly overlapping with the pump spot, is selected first. The only other mode that gets selected at a higher $P$ is the ground mode $E_0$ (blue). Thus, by modifying the cavity structure, we have isolated the effect of lasing-assisted ground-state condensation from the selection of further modes.

\begin{figure}[!htbp]
	\includegraphics{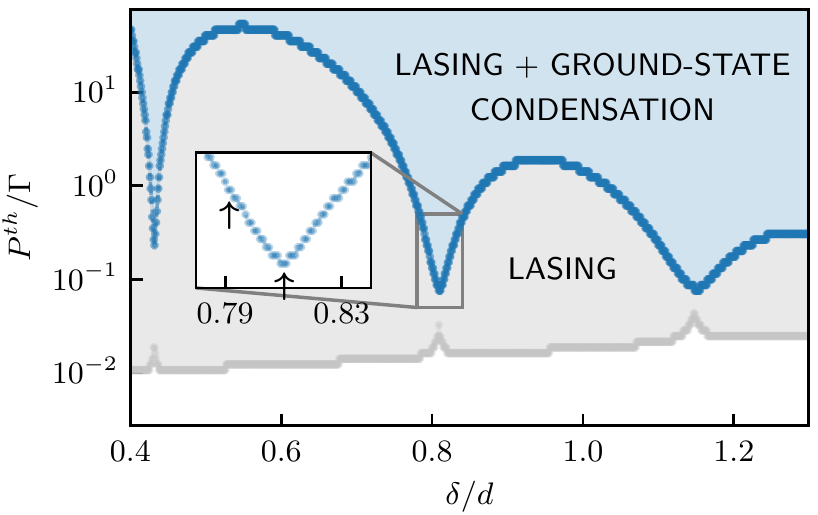}
	\caption{\label{fig:pd} Phase diagram showing the resonance effect on the ground-state condensate. The two phase boundaries are numerically obtained threshold pump rates $P^{th}$ for the lasing (gray) and the ground-state condensation (blue) as functions of $ \delta $. The arrows correspond to the case shown in Figs.~\ref{fig:modesDW} and \ref{fig:res}.}
\end{figure}

Figure.~\ref{fig:res}(b) shows the mode populations for the slightly larger parameter $\delta=0.81d$ [for which the double well potential essentially looks the same as the one depicted in Fig.~\ref{fig:modesDW}(a) for $\delta=0.79d $]. Note that this small parameter change leads to a large change in the separation between the first and the second threshold value. This strong sensitivity is caused by the delocalization of the lasing mode
(purple) over both wells [Fig.~\ref{fig:modesDW} (b)]. This is a result of the resonant coupling to a mode in the left well. As a result, the lasing-assisted creation of excited dye molecules in the left well is strongly enhanced and the second threshold to ground-state condensation happens at much lower pump rates. In 
Figure~\ref{fig:pd} we plot how the two threshold pump rates for lasing (gray curve) and ground-state condensation (blue curve) vary with $\delta$. The two arrows indicate the cases shown in Fig.~\ref{fig:res}. One can clearly observe a sequence of resonances at which the second threshold is strongly reduced. Remarkably, these resonances can be used to control the second threshold value by almost four orders of magnitude. In contrast, the threshold for the first transition shows merely small peaks at the resonances, which are associated with a reduced overlap with the pump spot due to delocalization. Thus, by engineering the transverse potential for the photons in the cavity, one can widely tune the separation between the first and the second threshold. 

%\paragraph{Conclusion}
We have shown that in a system of photons in a dye-filled cavity the interplay between driving (via gain and loss) and thermalization (via rovibrational relaxation of the dye molecules) can give rise to a robust mechanism for controlled two-mode emission. Namely, a transition to lasing in an excited cavity mode induced by an off-centered pump beam can trigger a second transition, where thermalization leads to the formation of a photon condensate in the ground mode. This mechanism can be made very robust and widely tuned by using a recently developed experimental technique for shaping the transverse potential for the photons in a trap.  

\acknowledgments
M.V. and A.E. acknowledge the support from the Deutsche Forschungsgemeinschaft (DFG) via the Research Unit FOR 2414 (under Project No. 277974659). H.A.M.L acknowledges financial support from the European Union FET-Open grant MIR-BOSE 737017 and from Provincia  Autonoma  di  Trento.

\end{document}